\documentclass[a4paper,10pt]{article}
\usepackage{hyperref}

\renewcommand{\title}[1]{ 
    \begin{center}%
    \Large\bf #1%
    \end{center}%
   }

\renewcommand{\author}[1]{%
    {\begin{center}
    #1
    \end{center}}}
\newcommand{\address}[1]{\vspace{-1.8em}\vspace{0pt}
    {\begin{center}
    \it #1 
    \end{center}}}

\begin{document}
\title{ IR   Finite Graviton Propagators  in de Sitter Spacetime  }
\author{ Mir Faizal\footnote{Email: mirfaizalmir@gmail.com}$\,^a$, Sudhaker Upadhyay\footnote{Email: 
sudhakerupadhyay@gmail.com} $^b$ and Bhabani Prasad Mandal\footnote{Email: bhabani.mandal@gmail.com} $^b$ }  
\address{ $\,^a${Department of Physics and Astronomy,    University of Lethbridge, \\  Lethbridge, 
AB T1K 3M4, Canada}}
\address{ $^b$ {Department of Physics, Banaras Hindu University,  \\
Varanasi-221005, India }}

 \begin{abstract}
Graviton propagator  diverges in certain gauges in de Sitter spacetime. We address this
problem in this work by   generalizing the infinitesimal BRST transformations in de Sitter spacetime to 
  finite field-dependent BRST (FFBRST) transformations. These FFBRST transformations are  symmetry of the classical action,
  but do not leave path integral measure invariant for  the graviton theory  in de Sitter spacetime. 
Due to the non-trivial Jacobian of such finite transformation the path integral measure 
changes and hence   FFBRST transformation is capable of relating theories 
in two different gauges. We explicitly construct FFBRST transformation 
which relates theory with diverging graviton  two-point function to theory with
infrared  (IR)   finite graviton. The FFBRST transformation thus establishes that     divergence 
 in graviton two-point function  may be only a gauge artifact. 
 \end{abstract}

\section{Introduction}
The  observations from   type I supernovae indicate that our universe has a positive cosmological constant and 
may approach de Sitter spacetime asymptotically \cite{super, super2, super0, super4, super5}. The de Sitter spacetime is also relevant in
the inflationary cosmology \cite{cos, cos0, cos1, cos2}. Inflaton fields 
corresponding to the open strings have been studied in brane-antibrane models \cite{z, za} and $D3/D7$ systems \cite{z2, z2a}, 
and the inflaton fields 
corresponding to the closed strings have been studied in Kahler moduli \cite{z4, z4a} and fibre inflation \cite{z7}. 
However, in all these models the realization of inflation depends crucially 
on the uplifting mechanism for de Sitter moduli stabilization \cite{coss}. This  uplifting mechanism occurs in  presence of 
$D3$-branes. 
It may be noted that even   the Wilson line approach  crucially depends 
on the uplifting mechanism for de Sitter moduli stabilization \cite{z5}. 
Due to relevance of de Sitter spacetime to inflation, it is important to study perturbative quantum gravity 
in de Sitter spacetime. 
However, graviton propagator in de Sitter spacetime found by Antoniadis et. al., suffered from IR   divergences \cite{1, 2}. 
In fact, these IR divergence   occur  in the covariant gauge for a certain  choices of gauge parameters, 
    $\beta = - n(n+3)/3$ with $n = 1, 2, 3 \cdots$ \cite{5}. 
However it is also possible to construct an IR   finite  graviton propagator 
  \cite{4, 4a, 4b}. So, there are strong indications to assume that the IR 
divergence that occur in the propagator by  Antoniadis et. al., are a gauge artifact.
This is supported from the fact  that  free  graviton propagator in covariant  gauge is equivalent to the IR finite
graviton   propagator    \cite{a}. However, in that analysis  
role of  interactions was not considered. What really needs to be demonstrated is that the generating functional for 
different values of the parameter $\beta$ are related to one another.  We  argue that the IR divergent  graviton propagators 
with $\beta = - n(n+3)/3$ are related to the  IR finite graviton propagators with 
 other values of $\beta$. However, 
to show that  explicitly,  
 we will need a formalism to connect the generating functionals 
 for the graviton propagators in the covariant gauge with different  values of the parameter $ \beta$.
 As the Euclidean approach has been used for calculation different propagators in de Sitter spacetime\cite{ix}, including the graviton 
 propagator \cite{xi}, we will also use 
 the Euclidean approach for calculating the graviton propagator. So, we will obtain the 
 function on a four dimensional sphere, and these Green's functions  are related to the 
 Feynman propagator in the de Sitter  spacetime through 
analytic continuation. It may be noted that we could have also used the planar patch of Lorentzian de Sitter for 
performing these calculations, however, the advantage of use the Euclidean approach is that it is easier to perform the FFBRST transformations 
in this approach.  We will use the   Euclidean vacuum  as the vacuum state
for performing these calculations \cite{va}.

The FFBRST transformation  \cite{bm}  was constructed systematically by integrating the usual BRST 
transformation \cite{masud}.  Such a generalized BRST transformations 
have the same form and properties of the usual BRST transformations except 
these do not leave path integral measure invariant. The non-trial Jacobian enables
such formulation to connect theories with different effective actions, hence FFBRST
transformations have found enormous number of applications in various branches
of high energy physics   \cite{bm, susk, jog, sb1,smm,rb, bm1,rbs1,sudd,sud,fs0,fs}. Similar generalization with same motivation 
and goal has also been done recently in slightly in different manner \cite{lav,ale,ale1}.
In this work we extend FFBRST formulation \cite{bm} in de Sitter spacetime and construct
appropriate finite field-dependent parameter to relate the generating functionals 
corresponding to effective theories with graviton propagator 
for various values of $\beta$.  It may be noted that even though we build this formalism motivated by the IR   divergences in   de Sitter spacetime, 
 this formalism is very general and can be used to relate generating functional for graviton propagator with 
 any arbitrary value of $\beta$. It may also be noted that there are real IR   divergences that occur in the ghost propagator 
 in de Sitter spacetime. However, modes responsible for these divergences 
do not contribute to loop diagrams in 
computations of  scattering amplitudes in perturbative quantum gravity and can thus be neglected \cite{b}.  It is possible to construct an effective IR 
finite ghost propagator for de Sitter spacetime utilizing the FFBRST transformation.
In this connection we would like to comment that the gaugeon formulation  \cite{yo0,SU,yok,yos,yoko}
which also connects different effective actions in perturbative 
quantum gravity \cite{6,7} could be other possibility to construct theories with IR finite 
graviton propagator. However gaugeon formalism has certain drawbacks that one need to
 introduce unphysical gaugeon fields in the theory and later 
 extra conditions are required to extract the physical states.

 It may be noted that there are various issues that 
are related the IR divergences in the graviton 
propagator.  
There are also several 
problems with the average gauges in de Sitter
space and 
any space with linearizion instabilities \cite{2}. 
It has also been argued that the main problem with 
certain values of gauge parameter is that for 
these values of the gauge parameter 
a   logarithmic divergences 
rather than power law divergences occurs \cite{w2,w1}. 
The power law divergences gets
automatically
subtracted  for
the allowed values of the gauge parameter.
In fact, it has been 
demonstrate using this line of argument, 
certain IR divergences also occur 
for  the allowed values of the gauge parameter
\cite{4, 4a}. Furthermore,   IR divergences 
which appear in certain gauges have the local
form of a gauge transformation,
but they need not be a symmetry
of the theory because
the needed “gauge” transformation 
diverges at infinity and
therefore invalidates the usual 
integration by parts and discarding
of surface terms is  needed to prove
invariance even of the classical
action \cite{1, q1, q2, p}. Even though we have 
neglected such terms in our paper by 
dropping a total divergence, however,
we would like to point out that there 
are many non-trivial issues relating to 
the occurrence of such divergences.
 It may be noted that even thought there are various 
 different sources of IR divergences, in this paper, 
 we will not address many of these issues. 
 We will rather demonstrate 
 that a graviton propagator in a certain gauge, 
 in which a certain kind of IR divergences 
 occurs can be related to the 
 graviton propagator in a different gauge where 
 such 
 IR divergences do not occur. 
 This can be done using the FFBRST transformations,
 as the FFBRST transformations are a symmetry 
 of the generating functional 
 and not of the effective  action
 which is obtained by adding the gauge fixing and ghosts terms to the 
 original action. In fact, it is this property of the 
 FFBRST transformation that has made it possible to use 
 the FFBRST transformation for analsying various 
 interesting physical systems \cite{bm, susk, jog, sb1,smm,rb, bm1,rbs1,sudd,sud,fs0,fs}. 
 Thus, motivated by such uses of the FFBRST transformations, we will analyse the 
 occurrence of a certain kind of IR divergences in this paper. 
 
 In this paper, we first study the perturbative quantum gravity on 
 curved space time where we   particularly emphasize on the de Sitter spacetime.
 The effective action of perturbative quantum gravity on de Sitter spacetime respects
 fermionic rigid BRST invariance. The BRST symmetry further generalize
 by making the parameter finite and field-dependent following the 
 techniques of Ref. \cite{bm}. 
 The FFBRST transformation generalized in such a way leads to
 a non-trivial Jacobian for functional measure.
 We show that for a particular choice of finite field-dependent parameter 
 the Jacobian relates the gauge parameters stimulating IR divergent and 
 IR finite graviton propagators. So, 
 in section 2, we analyse the perturbative quantum gravity in de Sitter spacetime, and in 
 section 3  we study the FFBRST transformation in de Sitter  spacetime. Then in section 4 we 
  relate  the   IR divergent  graviton two-point function to  
the  IR 
finite graviton propagators using the FFBRST transformations. In the final section we summaries the results.

\section{Perturbative quantum gravity}
 Let us first of all start by analysing the perturbative quantum gravity in de Sitter  spacetime. 
 The line element for de Sitter spacetime  which is a contracting and
expanding three-sphere is given by
\begin{equation}
ds^2=-dt^2+\frac{1}{H^2}\cosh^2(Ht)
[d\chi^2+\sin^2\chi(d\theta^2+\sin\theta^2d\phi^2)]\,, \label{metric}
\end{equation}
where $H$ is the Hubble constant.
In terms of  variable $\tau \equiv \pi/2-iHt$, the line element gets the following form:
\begin{equation}
ds^2=H^{-2}\left\{
d\tau^2+\sin^2\tau\,[d\chi^2+\sin^2\chi(d\theta^2+\sin\theta^2d\phi^2)]
\right\}\,,
\end{equation}
which is the line element of a four-dimensional sphere
of radius $H^{-1}$. Now we can set $H^2 = 1$. 
The Lagrangian density of pure gravity in  de Sitter spacetime  is given by,
\begin{equation}
{\cal L}_{ds} = \sqrt{- g^{(f)}}\,(R^{(f)} -6)\,, \label{fullL}
\end{equation} 
where we have set  $
 16 \pi G = 1 $. 
Here $g^{(f)}_{\mu\nu} $ is the  full metric and $R^{(f)}$ is the curvature corresponding to it. 
This  Lagrangian  is invariant under general coordinate transformations, 
\begin{equation}
\delta_\Lambda g^{(f)}_{\mu \nu} =  \pounds_\Lambda g^{(f)}_{\mu\nu},
\end{equation}
where  $\pounds_\Lambda g^{(f)}_{\mu\nu}  = 
\Lambda^\tau \nabla_\tau g^{(f)}_{\mu\nu} +g^{(f)}_{\mu \tau } \nabla_\nu \Lambda^\tau + g^{(f)}_{\mu \tau} \nabla_\mu \Lambda^\tau$,
is the Lie derivative. 
Now we can expand $g^{(f)}_{\mu\nu} $ in terms of a fixed background metric, $g_{\mu\nu}$, and a small perturbation around it, $h_{\mu\nu}$. 
\begin{equation}
 g^{(f)}_{\mu\nu} = g_{\mu\nu} + h_{\mu\nu}.
\end{equation}
Now we can also expand the Lagrangian for gravity with a cosmological constant in terms of this fixed 
background metric and a small perturbation around it. Furthermore, this small perturbation is regarded as the quantum field 
to be quantized in perturbative quantum gravity. It may be noted that this Lagrangian will contain infinitely many terms
because the original Lagrangian contained the inverse of the metric in it. 
Now as $g_{\mu\nu}$ is fixed, the transformation of $g^{(f)}_{\mu\nu}$ will be attributed to $h_{\mu\nu}$,
$\delta_\Lambda h_{\mu\nu} = \pounds_\Lambda g^{(f)}_{\mu\nu} = 
  \pounds_\Lambda g_{\mu\nu} +   \pounds_\Lambda h_{\mu\nu} $. 
Now to the first order in $\Lambda_\mu$, the  Lagrangian  for perturbative quantum gravity will be invariant 
 to all orders in $h_{\mu\nu}$, under the following transformation, 
\begin{equation}
\delta_\Lambda h_{\mu\nu} =
\nabla_\mu \Lambda_\nu + \nabla_\nu \Lambda_\mu + \pounds_\Lambda h_{\mu\nu},\label{gauge}
\end{equation}
where the Lie derivative $\pounds_\Lambda h_{\mu\nu}$ is given by
$ \pounds_\Lambda h_{\mu\nu} = \Lambda^\tau \nabla_\tau h_{\mu\nu} +h_{\mu \tau } \nabla_\nu \Lambda^\tau + h_{\nu\tau} \nabla_\mu \Lambda^\tau
$.   
The resulting Lagrangian density for the linearized gravity   is 
written, after
dropping a total divergence, as  
\begin {eqnarray} 
\mathcal{L}_{ds} &=& \mathcal{L}_{f} + \mathcal{L}_{int} \nonumber \\ &=& \sqrt{-g}
\left[ \frac{1}{2}\nabla_{\mu}h^{\mu \eta}\nabla^{\nu}h_{\nu \eta}
-\frac{1}{4}\nabla_{\mu}h_{\nu \eta}\nabla^{\mu}h^{\nu \eta}
+\frac{1}{4}(\nabla^{\mu}h-2\nabla^{\nu} h^{\mu}_{\ \nu})\nabla_{\mu}h \right.
 \nonumber \\
&& \left. \ \ \ \ \ \ 
-\frac{1}{2}\left(
h_{\mu \nu}h^{\mu \nu}+\frac{1}{2}h^2\right)\right] + \mathcal{L}_{int}, \label{Lagden}
\end{eqnarray}
with $h = h^{\mu}_{\ \mu}$ and $\mathcal{L}_{int}$ is the interaction part of the Lagrangian.

To quantize the theory we need to break this gauge invariance
for canonical quantization.
Here this is achieved by choosing a general (covariant) gauge fixing condition for this 
Lagrangian   as, 
\begin{equation}
 [\nabla^\nu h_{\mu \nu} - k \nabla_\mu  h] =0, 
\end{equation}
where $k \neq 1$. It may be noted that for $k =1$, the gauge redundancies are not fully removed, and so, usually $k$ is written as 
$1 + \beta^{-1}$ for a finite value of $\beta$. The gauge fixing condition can be incorporate at a quantum level by addition a 
gauge 
fixing term to the original Lagrangian, 
\begin{equation}
 \mathcal{L}_{gf} = \sqrt{-g} b^\mu [\nabla^\nu h_{\mu \nu} - (1 + \beta^{-1}) \nabla_\mu  h]  +\sqrt{-g} \frac{\alpha}{2} b^\mu b_\mu. 
\end{equation}
We can obtain the ghost term corresponding to this gauge fixing term, by first taking the gauge transformation 
of the gauge fixing condition,  then replacing all the gauge parameters with ghosts, and finally contracting 
the quantity thus obtained with anti-ghosts. 
Thus, the ghost term corresponding to this gauge fixing term  can be written as 
\begin{eqnarray}
  \mathcal{L}_{gh} &=&\sqrt{-g} \bar c^\mu \nabla^\nu [ \nabla_\mu c_\nu + \nabla_\nu c_\mu - 2 (1 + \beta^{-1}) g_{\mu\nu} \nabla^\tau c_\tau +
  ( \pounds_c h_{\mu\nu} \nonumber \\ && - (1 + \beta^{-1}) g_{\mu\nu}g^{\tau \rho} \pounds_c h_{\tau \rho} ) ], 
\end{eqnarray}
where $\pounds_c h_{\mu\nu}$ is given by $\pounds_c h_{\mu\nu} = c^c \nabla_c h_{\mu\nu} +h_{\mu \sigma } \nabla_\nu c^\sigma 
+ h_{\nu\sigma} \nabla_\mu c^\sigma $. Now the sum of the deformed  Lagrangian for gravity, the gauge fixing term and the ghost term 
is invariant under the following BRST transformations, 
\begin{eqnarray}
 s b_\mu = 0, &&  s h_{\mu\nu} = \nabla_\mu c_\nu + \nabla_\nu c_\mu +    \pounds_c h_{\mu\nu},  \nonumber\\
 s \bar c^\mu = b^\mu,  && s c_\mu  =   c^\nu \nabla_\nu c_\mu.  \label{brs}
\end{eqnarray}
It may be noted that the invariance of the sum of the non-local Lagrangian for gravity, along with the gauge fixing term 
and the  ghost term under the BRST transformation follows from the nilpotency of the BRST transformations, 
  $ s\mathcal{L} = 0$. 
This is because the sum of the gauge fixing term and the ghost term can be written  as a total BRST variation, 
\begin{equation}
  \mathcal{L}_{gf} + \mathcal{L}_{gh} = s \left[\sqrt{-g}\bar c^\mu \left[\nabla^\mu h_{\mu \nu} - (1 + \beta^{-1}) \nabla_\mu h 
  +   \frac{\alpha}{2} b_\mu \right]\right]. 
\end{equation}
It is known that the graviton two-point function is IR   diverges for $\beta = -n(n +3)/3$ \cite{5}. 
So, we  take the initial gauge fixing condition  as 
\begin{equation}
 G_1[h]= \left[\nabla^\nu h_{\mu \nu} - (1 + \beta_1)^{-1} \nabla_\mu  h   \right] =0, 
\end{equation}
where 
\begin{equation}
\beta_1 =  - \frac{n(n+3)}{3} +  \epsilon, 
\end{equation}
and we take the limit  $\epsilon \to 0$, at the end of our calculation. 
In the next section, we demonstrate  that it is possible to gauge away this IR   divergence using the FFBRST transformation. 
Hence,  an IR  finite propagator is obtained  even after taking the limit. These IR   divergences are only 
gauge artifacts. 
We take our final gauge fixing condition to be 
\begin{equation}
 G_2[h]=  \left[\nabla^\nu h_{\mu \nu} - (1 + \beta_2)^{-1} \nabla_\mu  h\right] =0, 
\end{equation}
where 
\begin{equation}
\beta_2 \neq  - \frac{n(n+3)}{3}.
\end{equation}
Now we can write the sum of the gauge fixing and ghost terms for the initial  gauge fixing condition as 
\begin{equation}
   \mathcal{L}_{1gf} + \mathcal{L}_{1gh}= s \left[\sqrt{-g}\bar c^\mu \left[\nabla^\mu h_{\mu \nu} - (1 + \beta_1^{-1}) \nabla_\mu h 
  +   \frac{\alpha}{2} b_\mu \right]\right], 
\end{equation}
and we can also write the sum of the gauge fixing and ghost terms for the final gauge fixing condition as 
\begin{equation}
     \mathcal{L}_{2gf} + \mathcal{L}_{2gh} =s \left[\sqrt{-g}\bar c^\mu \left[\nabla^\mu h_{\mu \nu} - (1 + \beta_2^{-1}) \nabla_\mu h 
  +   \frac{\alpha}{2} b_\mu \right]\right]. 
\end{equation}
Now we can write the total Lagrangian as the sum of the original Lagrangian, the gauge fixing and ghost terms. 
\begin{eqnarray}
 \mathcal{L}_1 &=& \mathcal{L}_{ds} +   \mathcal{L}_{1gf} + \mathcal{L}_{1gh},   \nonumber \\
  \mathcal{L}_2 &=& \mathcal{L}_{ds} +   \mathcal{L}_{2gf} + \mathcal{L}_{2gh}. 
\end{eqnarray}
We can neglect the interaction part of the Lagrangian and write the equation of motions, from these Lagrangian's as follows, 
\begin{eqnarray}
 L_{\mu \nu(\beta_1)}^{\sigma \lambda } h_{\sigma \lambda  } =0,  \nonumber \\
  L_{\mu \nu (\beta_2) }^{\sigma \lambda  } h_{\sigma \lambda  } =0, 
\end{eqnarray}
where 
 \begin{eqnarray}
 L_{\mu \nu (\beta_1) }^{\sigma \lambda} h_{\sigma \lambda  }  &=& \frac{1}{2} \nabla^\tau \nabla_\tau h_{\mu\nu} 
 - \left(\frac{1}{2} - \frac{1}{2 \alpha}\right) 
 (\nabla_\mu \nabla_\tau h^\tau_\nu + \nabla_\nu \nabla_\tau h^\tau_\mu )
    \nonumber \\ && + \left( \frac{1}{2}  - \frac{\beta_1 +1 }{ \alpha \beta_1} \right)\nabla_\mu \nabla_\nu  h 
   + \left( \frac{(\beta_1 + 1)^2}{\alpha \beta_1^2}  - \frac{1}{2}\right) g_{\mu\nu} \nabla^\tau \nabla_\tau h 
    \nonumber \\ &&  +\left(\frac{1}{2 }  - \frac{1 + \beta_1}{\alpha \beta_1}\right)  g_{\mu \nu}\nabla^\tau \nabla^\rho h_{\tau \rho}
    - h_{\mu \nu} - \frac{1}{2} g_{\mu \nu }h,  \nonumber \\
 L_{\mu \nu (\beta_2) }^{\sigma \lambda} h_{\sigma \lambda  }  &=& \frac{1}{2} \nabla^\tau \nabla_\tau h_{\mu\nu} 
 - \left(\frac{1}{2} - \frac{1}{2 \alpha}\right) 
 (\nabla_\mu \nabla_\tau h^\tau_\nu + \nabla_\nu \nabla_\tau h^\tau_\mu )
   \nonumber \\&& + \left( \frac{1}{2} - \frac{\beta_2+1 }{ \alpha \beta_2} \right)\nabla_\mu \nabla_\nu   h 
  + \left( \frac{(\beta_2 + 1)^2}{\alpha \beta_2^2}  - \frac{1}{2}\right) g_{\mu\nu} \nabla^\tau \nabla_\tau  h 
     \nonumber \\ &&  +\left(\frac{1}{2 }  - \frac{1 + \beta_2}{\alpha \beta_2}\right)  g_{\mu \nu}\nabla^\tau \nabla^\rho h_{\tau \rho}
    - h_{\mu \nu} -\frac{1}{2} g_{\mu \nu}h. 
\end{eqnarray}
So, we can write the graviton two-point function for the two different gauges can be defined as 
\begin{eqnarray}
 L_{\mu\nu (\beta_1)}^{\tau  \rho } G^{(\beta_1)}_{\tau \rho \tau' \rho' } (x, x') &=& \delta_{\mu \nu \tau' \rho'}(x, x'), 
 \nonumber \\
  L_{\mu\nu (\beta_2)}^{\tau  \rho } G^{(\beta_2)}_{\tau \rho \tau' \rho' } (x, x') &=& \delta_{\mu \nu \tau' \rho'}(x, x'). 
\end{eqnarray}
This graviton two-point function  in the final  gauge is IR   finite  \cite{4, 4a, 4b}, 
where as the graviton two-point function in the initial gauge
diverges for $\epsilon \to 0$ \cite{5}. It may be noted that the exact expression for the graviton propagator will depend on the 
exact choice of the gauge, however, the important  point is that such a graviton propagator is IR finite in a certain gauge. 
 We now will proceed to show that it is possible to construct 
 a particular FFBRST transformation which takes the generating functional 
 in initial gauge to 
 that of in final gauge. Thus the FFBRST takes the theory with IR divergent 
 graviton propagator to the theory with IR finite graviton propagator. This indicates IR 
 divergence of graviton propagator is an gauge artifact.
 It may be noted that even though we have performed our analysis to relate the generating functionals in the 
 de Sitter spacetime, this analysis is very general and can be used to relate generating functionals with   
 with different values of $\beta$ in  any curved spacetime. 
 \section{FFBRST transformation}
 To construct the FFBRST transformation \cite{bm} for theory of quantum gravity in de Sitter spacetime we first
write the 
usual BRST transformation characterized by an infinitesimal 
Grassmann parameter  $\delta\Lambda$ as follows, 
\begin{equation} 
\delta_b \Phi^i  (x) =s\Phi^i (x) \delta \Lambda,
\end{equation} 
where $\Phi^i (x) =  (h^{\mu\nu} (x), c^\mu (x), \bar c^\mu (x), b^\mu (x))$ is the set of all fields in the theory. 
It may be noted  that the invariance under BRST transformation is not affected by 
  $\delta\Lambda$  being (i) finite or infinitesimal, (ii) field-dependent or field-independent, as long 
as it is anticommuting and spacetime independent.
This motivates  the  generalization of the BRST transformations to finite field dependent BRST transformations. 
This is done by first making the infinitesimal parameter $\delta \Lambda$ field dependent, and introducing an arbitrary 
parameter $\kappa$, such that $ 0 \leq  \kappa \leq 1   $. Then all the fields are made to depend on $\kappa$, in such a way that 
$ \Phi^i (x, \kappa = 0)  $ are the initial fields, and $ \Phi^i (x, \kappa =1) $ are the transformed fields. 
Now we can write the field dependent infinitesimal  BRST transformations as, 
\begin{eqnarray}
s \,h_{\mu\nu} &=& (\nabla_\mu c_\nu + \nabla_\nu c_\mu + \pounds_{(c)} h_{\mu\nu})\ \Theta'[\Phi (x, \kappa) ], \nonumber \\
s \,c^\mu &=&   c_\nu \nabla^\nu c^\mu\ \Theta'[\Phi (x, \kappa) ], \nonumber \\
s \,\bar{c}^\mu &=& b^\mu \ \Theta'[\Phi (x, \kappa)], \nonumber \\ 
s \,b^\mu  &=& 0,\label{ff}
\end{eqnarray}
where $\Theta ' [\Phi (x, \kappa)]$ is infinitesimal but field dependent  parameter. 
Now the finite field dependent BRST  (FFBRST) transformations, can be constructed by integrating such infinitesimal 
field dependent BRST transformations   \cite{6}, 
\begin{eqnarray}
f \,h_{\mu\nu} &=& (\nabla_\mu c_\nu + \nabla_\nu c_\mu + \pounds_{(c)} h_{\mu\nu})\ \Theta[\Phi (x) ], \nonumber \\
f \,c^\mu &=&   c_\nu \nabla^\nu c^\mu\ \Theta[\Phi (x) ], \nonumber \\
f \,\bar{c}^\mu &=& b^\mu \ \Theta[\Phi (x) ], \nonumber \\ 
f \,b^\mu  &=& 0,\label{ff}
\end{eqnarray}
where the finite field dependent parameter is given by 
\begin{equation}
\Theta [ \Phi (x)]  = \int_0^1 d\kappa \Theta' [\Phi (x, \kappa)]. 
\end{equation}
So, the FFBRST transformation can be written as  
\begin{equation}
f \Phi^i (x)= \Phi^i (x,\kappa =1)-\Phi^i(x,\kappa=0)= s [\Phi^i(x) ]\Theta[\Phi  ]. 
\label{kdep}
\end{equation}

 This FFBRST  transformation  is  symmetry  of the effective action only but not of the 
 functional measure  because the
path integral measure changes under FFBRST transformation to local functional of fields.  
Suppose the Jacobian of the path integral measure under such transformation is  written  by
\begin{eqnarray}
{\cal D}\Phi  &=&J(\kappa) {\cal D}\Phi (\kappa),\nonumber\\
  &=& J(\kappa +d\kappa){\cal D}\Phi (\kappa +d\kappa).\label{jac1}
\end{eqnarray}
The transformation from $\Phi (\kappa)$ to $\Phi (\kappa +d\kappa)$ is an infinitesimal one and one has, for 
its Jacobian
\begin{equation}
\frac{J(\kappa)}{J(\kappa +d\kappa)}=\int d^4x\sum_{\Phi }\pm\frac{\delta{\Phi^i (x,\kappa +d\kappa)}}{\delta{\Phi^i(x,\kappa)}},
\end{equation}
where $\sum_{\Phi }$ sums over all the fields in the measure 
 and the $\pm$ sign refers to the cases of fields $\Phi^i (x, \kappa)$ being
bosonic or fermionic in nature. 
Now utilizing Taylor's expansion the infinitesimal change in the Jacobian  
is calculated by \cite{6}
\begin{equation}
\frac{1}{J}\frac{dJ}{d\kappa}=-\int d^4y \sum_{\Phi } \left [\pm s\Phi^i (y,\kappa )\frac{
\delta\Theta^\prime [\Phi ]}{\delta\Phi^i (y,\kappa )}\right].\label{jac}
\end{equation}
 Now the Jacobian, $J(\kappa )$, can be replaced   (within the functional integral)  by
\begin{equation}
J(\kappa )\rightarrow \exp[iS_1[\Phi (x,\kappa), \kappa ]],\label{s}
\end{equation}
where $ S_1[\Phi  (x), \kappa]$ is local functional of fields, if 
 the following condition is satisfied   
 \begin{eqnarray}
 \frac{1}{J}\frac{dJ}{d\kappa}-i\frac
{dS_1[\Phi (x,\kappa ), \kappa]}{d\kappa}    =0. \label{mcond}
\end{eqnarray}
Therefore, by constructing an appropriate $\Theta'$, we are able calculate the non-trivial (local) Jacobian
 which extends the effective action by a term $S_1$.

 \section{Recovering IR  finite the graviton propagators}
 In the previous  section, we have analysed the FFBRST transformation for the de Sitter spacetime in a general gauge. 
 In this section, we explicitly use the results of the previous section to demonstrate that the 
 IR   divergence in the graviton two-point function can be gauged away. 
 The FFBRST transformation   for perturbative quantum gravity   corresponding to the BRST transformation  
are written as given by Eq. (\ref{ff}). Now to show the connection
between IR  divergent the graviton propagators and IR  finite the graviton propagators
we derive specific $\Theta[\Phi ]$ constructed from following infinitesimal field-dependent parameter: 
 \begin{eqnarray}
 \Theta'[\Phi ] =\int d^4x \ \sqrt{-g}\bar c^\mu\left[(\beta_1^{-1} -\beta_2^{-1})\nabla_\mu h   \right].
 \label{the}
 \end{eqnarray}
 Using Eq. (\ref{jac}) along with  Eq. (\ref{the}),  we calculate
 the infinitesimal change in Jacobian as follows,
 \begin{eqnarray}
 \frac{1}{J(\kappa)}\frac{dJ(\kappa)}{d\kappa}&=&\int d^4x \sqrt{-g}\left[ -b^\mu(-\beta_1^{-1} +\beta_2^{-1})\nabla_\mu h  
 -\bar c^\mu \nabla^\nu \left[ 2(-\beta_1^{-1} +\beta_2^{-1}) g_{\mu\nu}\nabla^\tau c_\tau 
\right.\right. \nonumber\\ 
&& + \left.\left. (-\beta_1^{-1} +\beta_2^{-1})g_{\mu\nu}g^{\tau\rho} \pounds_c h_{\tau\rho} \right]\right].\label{jac1}
 \end{eqnarray}
 Now we make an ansatz for local functional $S_1[\Phi ,\kappa]$, which can be written as  
 \begin{eqnarray}
 S_1[\Phi ,\kappa]&=&\int d^4x \sqrt{-g}\left[ \xi_1(\kappa) b^\mu(\beta_1^{-1} )\nabla_\mu h  + \xi_2(\kappa)b^\mu(\beta_2^{-1})\nabla_\mu h  
 +\xi_3(\kappa)\bar c^\mu \nabla^\nu (\beta_1^{-1}) g_{\mu\nu}\nabla^\tau c_\tau 
\right.\nonumber\\ 
&& + \left. \xi_4(\kappa)\bar c^\mu \nabla^\nu (\beta_2^{-1}) g_{\mu\nu}\nabla^\tau c_\tau 
+\xi_5(\kappa)\bar c^\mu \nabla^\nu(\beta_1^{-1} )g_{\mu\nu}g^{\tau\rho} \pounds_c h_{\tau\rho} \right.\nonumber\\ 
&& + \left. \xi_6(\kappa)\bar c^\mu \nabla^\nu(\beta_2^{-1} )g_{\mu\nu}g^{\tau\rho} \pounds_c h_{\tau\rho} \right],\label{s1}
 \end{eqnarray}
 where $\xi_i (i=1,2,..,6)$ are arbitrary constant $\kappa$-dependent parameters
 which can be evaluated from the 
 essential condition given by  Eq.  (\ref{mcond}).
 The essential condition given by Eq. (\ref{mcond}) along  with Eq.  (\ref{jac1}) and  Eq. (\ref{s1}), 
 leads to following differential equations:
 \begin{eqnarray}
 \xi_1'-1=0, &\xi_2'+1=0, &  \xi_3'-2=0, \nonumber \\
 \xi_4'+2=0, & \xi_5'-1=0, & \xi_6'+1=0.
 \end{eqnarray}
 The particular solutions of above equations satisfying initial boundary 
 conditions $\xi_i (\kappa=0)=0$ are
 \begin{eqnarray}
 \xi_1 = 1, & \xi_2 = -1, & \xi_3 =2,\nonumber \\ \xi_4 =-2, & \xi_5 =1, & \xi_6 =-1.
 \end{eqnarray}
 With this values of parameters the local functional comprising Jacobian 
 is given by
 \begin{eqnarray}
 S_1[\Phi ,\kappa]&=&\int d^4x \sqrt{-g}\left[ \kappa b^\mu(\beta_1^{-1} )\nabla_\mu h  -\kappa b^\mu(\beta_2^{-1})\nabla_\mu h  
 +2\kappa\bar c^\mu \nabla^\nu (\beta_1^{-1}) g_{\mu\nu}\nabla^\tau c_\tau 
\right.\nonumber\\ 
&&-\left. 2\kappa\bar c^\mu \nabla^\nu (\beta_2^{-1}) g_{\mu\nu}\nabla^\tau c_\tau 
+\kappa\bar c^\mu \nabla^\nu(\beta_1^{-1} )g_{\mu\nu}g^{\tau\rho} \pounds_c h_{\tau\rho} \right.\nonumber\\ 
&&- \left. \kappa\bar c^\mu \nabla^\nu(\beta_2^{-1} )g_{\mu\nu}g^{\tau\rho} \pounds_c h_{\tau\rho} \right],
 \end{eqnarray}
 which vanishes at $\kappa=0$.  However, at $\kappa=1$ it translates to
  \begin{eqnarray}
 S_1[\Phi ,1]&=&\int d^4x \sqrt{-g}\left[ b^\mu(\beta_1^{-1} )\nabla_\mu h  - b^\mu(\beta_2^{-1})\nabla_\mu h  
 +2\bar c^\mu \nabla^\nu (\beta_1^{-1}) g_{\mu\nu}\nabla^\tau c_\tau 
\right.\nonumber\\ 
&&-\left. 2\bar c^\mu \nabla^\nu (\beta_2^{-1}) g_{\mu\nu}\nabla^\tau c_\tau 
+\bar c^\mu \nabla^\nu(\beta_1^{-1} )g_{\mu\nu}g^{\tau\rho} \pounds_c h_{\tau\rho} \right.\nonumber\\ 
&&-\left. \bar c^\mu \nabla^\nu(\beta_2^{-1} )g_{\mu\nu}g^{\tau\rho} \pounds_c h_{\tau\rho} \right],
 \end{eqnarray}
 which implies that the  Jacobian can be written as  $\exp{iS_1[\Phi,1]}$. Now, this Jacobian
 changes the effective action of the path integral as follows,
 \begin{equation}
  S_{1gf} + S_{1gh} + S_1[\Phi ,1] =   S_{2gf} + S_{2gh}, 
 \end{equation}
where 
\begin{eqnarray}
 S_{1gf} + S_{1gh} &=&  \int d^4x\sqrt{-g}\left( b^\mu [\nabla^\nu h_{\mu \nu} - (1 + \beta_1^{-1}) \nabla_\mu  h]  \right. \nonumber \\
   &&\left.  + \frac{\alpha}{2} b^\mu b_\mu
+ \bar c^\mu \nabla^\nu [ \nabla_\mu c_\nu + \nabla_\nu c_\mu - 2 (1 + \beta_1^{-1}) g_{\mu\nu} \nabla^\tau c_\tau \right. \nonumber \\
   &&\left. +
  ( \pounds_c h_{\mu\nu} - (1 + \beta_1^{-1}) g_{\mu\nu}g^{\tau \rho} \pounds_c h_{\tau \rho} ) 
  ]\right), 
 \nonumber \\
  S_{2gf} + S_{2gh} &=&\int d^4x\sqrt{-g}\left( b^\mu [\nabla^\nu h_{\mu \nu} - (1 + \beta_2^{-1}) \nabla_\mu  
  h] \right. \nonumber \\
   &&\left. + \frac{\alpha}{2} b^\mu b_\mu + \bar c^\mu \nabla^\nu [ \nabla_\mu c_\nu + \nabla_\nu 
   c_\mu - 2 (1 + \beta_2^{-1}) g_{\mu\nu} \nabla^\tau c_\tau \right. \nonumber \\
   &&\left. +
  ( \pounds_c h_{\mu\nu} - (1 + \beta_2^{-1}) g_{\mu\nu}g^{\tau \rho} \pounds_c h_{\tau \rho} ) 
  ]\right).
\end{eqnarray}
We can write the action sum of the original classical action, the gauge fixing term, and the ghost term for 
initial gauge as 
\begin{equation}
 S_{1T} = \int d^4 x   [\mathcal{L}_{ds} +    \mathcal{L}_{1gf} + \mathcal{L}_{1gh}], 
\end{equation}
and   we can write the action sum of the original classical action, the gauge fixing term, and the ghost term for 
final  gauge as 
\begin{equation}
 S_{2T} = \int d^4 x  [\mathcal{L}_{ds}  +    \mathcal{L}_{2gf} + \mathcal{L}_{2gh}]. 
\end{equation}
 Now we can take the limit $\epsilon \to 0$ for this transformed action and obtain 
  \begin{equation}
 \lim_{\epsilon \to 0} S_{1T }  +  \lim_{\epsilon \to 0}  S_1[\Phi ,1] =    \lim_{\epsilon \to 0}  S_{2T}.
 \end{equation}
 The two-point function obtained from the action  $ S_{1T }$ is IR   
 divergence in the limit $\epsilon  \to 0$, and the 
 two-point function obtained from the action   $ S_{2T }$ is IR    finite in the limit $\epsilon  \to 0$. 
 Hence, in the limit  $\epsilon  \to 0$, the action $ S_1[\Phi ,1] $ also produces IR   divergent contributions and these contributions 
 exactly cancel the IR   divergence coming from $ S_{1T }$ as follows 
 \begin{eqnarray}
 \lim_{\epsilon \to 0} \int {\cal D}\Phi   {\cal O}[\Phi]_2   e^{i S_{1T}[\Phi ] }
 \stackrel{FFBRST}{---\longrightarrow}\lim_{\epsilon \to 0} \int {\cal D}\Phi   {\cal O}[\Phi]_2   e^{i S_{2T}[\Phi ] },
\end{eqnarray} 
where $ {\cal O}[\Phi]_2 $ is two-point composite operator. We can neglect the interactions, and calculate the relation between Green's functions. 
Now if $G_{\mu\nu\tau'\rho'} (x, x')$ is the contribution to the Green's function coming from $ S_1[\Phi ,1]$, then we can write  
\begin{equation}
\lim_{\epsilon \to 0}\left[G_{\mu\nu\tau'\rho'} (x, x') + G^{(\beta_1)}_{\mu\nu\tau'\rho'} (x, x') \right] = 
\lim_{\epsilon \to 0}\left[  G^{(\beta_2)}_{\mu\nu\tau'\rho'} (x, x') \right].
\end{equation}
 Hence, it has been possible to gauge away these IR   divergences in the graviton two-point function in de  Sitter spacetime using 
 FFBRST transformation. This implies that these IR   divergences are only gauge artifacts. It may be noted as the gauge fixing and the ghost terms 
 are not effected by including interactions in the action. In fact, it has been demonstrated that the FFBRST transformation are a symmetry 
 of the generating functional, and so it would be possible to relate the two generating functional even after the interactions have been added. 
 It may be noted as the FFBRST transformations relate the full generating functional, they have  been applied to various interesting 
 physical systems \cite{bm, susk, jog, sb1,smm,rb, bm1,rbs1,sudd,sud,fs0,fs}. Thus,  even if  we do not neglect the interactions,  we have 
\begin{eqnarray}
 S_{IT} &=& S_{f} + S_{int} + S_{1gf} + S_{1gh}, \nonumber \\
 S_{2T} &=& S_{f} + S_{int}+  S_{2gf} + S_{2gh},
\end{eqnarray}
where $S_{int}$ are the interactions in the perturbative quantum gravity.
So,  we can take the limit $\epsilon \to 0$ for the actions, even after taking the interactions into account,  
  \begin{equation}
 \lim_{\epsilon \to 0} \left[S_{1T }  +     S_1[\Phi ,1]\right] =    \lim_{\epsilon \to 0}  S_{2T}.
 \end{equation}
 Now for any operator ${\cal O}[\Phi]$, we can formally write  
   \begin{eqnarray}
 \lim_{\epsilon \to 0} \int {\cal D}\Phi   {\cal O}[\Phi]   e^{i S_{1T}[\Phi ] }
 \stackrel{FFBRST}{---\longrightarrow}\lim_{\epsilon \to 0} \int {\cal D}\Phi   {\cal O}[\Phi]   e^{i S_{2T}[\Phi ] }. 
\end{eqnarray} 
 Hence, at least formally we can argue that such IR divergence will not occur even in loop calculations. 
 However, as we were interested in demonstrating the 
 relation between two Green's functions in this paper, we have explicitly only demonstrated that for the Green's functions. 
 \section{Conclusion}
  In this paper, we have analysed perturbative quantum gravity on de Sitter spacetime. The BRST and FFBRST 
  transformations for the  perturbative quantum gravity were explicitly constructed in de Sitter 
  spacetime. The FFBRST transformations were used to relate the generating functionals with different 
  values of the parameter $\beta$. We construct appropriate finite field-dependent parameter
  such that the Jacobian contribution of path integral measure relates the   graviton propagator with  a IR divergence  to the IR finite 
  graviton propagator. Thus, it was argued that it might be possible that certain kind of IR  divergence in the graviton propagator are 
  only a gauge artifact. However, we would like to point out that there are arguments  try to argue  that the  
  removal of such divergences is only an artifact of the regularization procedure \cite{w2,w1}. 
 Since the spacetime noncommutativity changes the IR behavior of   quantum field theories  \cite{1aaaa, 2aaaa}, 
 and perturbative quantum gravity has been studied on noncommutative spacetime \cite{pqg1, pqg2, pqg3, pqg4, pqg4a},   
 it would also be interesting analyse the IR divergence's in de Sitter  spacetime 
 in noncommutative spacetime.


\begin{thebibliography}{100}
\bibitem{super}A.G. Riess et al., Astron. J. 116, 1009 (1998)
\bibitem{super1}S. Perlmutter et al., Nature 391 , 51 (1998) 
\bibitem{super2}A. G. Riess et al., Astron. J. 118, 2668 (1999)
\bibitem{super0}S. Perlmutter et al., Astrophys. J. 517, 565 (1999)
\bibitem{super4}A. G. Riess et al., Astrophys. J. 560, 49 (2001)
\bibitem{super5} J. L. Tonry et al., Astrophys. J. 594, 1 (2003)
\bibitem{cos} K.  Bamba, G. Cognola, S. D. Odintsov and S.  Zerbini,  Phys. Rev. D 90, 023525 (2014) 
\bibitem{cos0} A.  del Rio and J.  Navarro-Salas,  Phys. Rev. D89, 084037 (2014)  
\bibitem{cos1} D.  Seery,  JCAP.  0905, 021 (2009)
\bibitem{cos2} S. Dubovsky, L. Senatore and G.  Villadoro,   JHEP.  0904, 118 (2009) 
\bibitem{z} C. P. Burgess, M. Majumdar, D. Nolte, F. Quevedo, G. Rajesh and R. J. Zhang,   JHEP.  0107, 047 (2001)
\bibitem{za} S.  Kachru, R.  Kallosh, A.  Linde, J.  Maldacena, L.  McAllister and S. P. Trivedi,   JCAP.  0310, 013 (2003)
\bibitem{z2} K.  Dasgupta, C.  Herdeiro, S.  Hirano and R.  Kallosh, Phys. Rev. D65, 126002 (2002)
\bibitem{z2a} C. P. Burgess, J. M. Cline and M. Postma, JHEP.  0903, 058 (2009)
\bibitem{z4} J. P. Conlon and F.  Quevedo, JHEP.  0601, 146 (2006)
\bibitem{z4a} J. R. Bond, L. Kofman, S. Prokushkin and P. M. Vaudrevange, Phys. Rev. D75, 123511 (2007)
\bibitem{z7} M. Cicoli, C. P. Burgess and F. Quevedo, JCAP.  0903, 013 (2009)
\bibitem{coss}S.  Krippendorf and F.  Quevedo,  JHEP.  0911, 039 (2009) 
\bibitem{z5} A. Avgoustidis, D. Cremades and F. Quevedo, Gen. Rel. Grav. 39, 1203 (2007)
\bibitem{1} I. Antoniadis, J. Iliopoulos, and T. N. Tomaras, Phys. Rev. Lett. 56, 1319 (1986)
\bibitem{2} I. Antoniadis and E. Mottola, J. Math. Phys. 32, 1037 (1991)
\bibitem{5} B. Allen, Phys. Rev. D 34, 3670 (1986)
\bibitem{4} A. Higuchi and R. H. Weeks, Class. Quant. Grav. 20, 3005 (2003)
\bibitem{4a} A.  Higuchi, D.  Marolf and I. A. Morrison,  Class. Quant. Grav. 28, 245012 (2011) 
\bibitem{4b} R. P. Bernar, L.  C. B. Crispino and A.  Higuchi, Phys. Rev. D90, 024045 (2014) 
\bibitem{a}M. Faizal and A. Higuchi, Phys. Rev. D85,  12402  (2012) 
\bibitem{ix} B. Allen and T. Jacobson, Commun. Math. Phys. 103, 669 (1986)
\bibitem{xi}A. Higuchi and S. S. Kouris, Class. Quantum Grav. 18, 4317 (2001)
\bibitem{va}  G. Gibbons and S. W. Hawking, Phys. Rev. D15, 2738 (1977)
\bibitem{bm} S. D. Joglekar and B. P. Mandal, Phys. Rev. D 51, 1919 (1995) 
 \bibitem{masud}M. Chaichian and  N. F. Nelipa,  \textit{Introduction to Gauge Field Theories},
Springer Berlin Heidelberg, (2012)
  \bibitem{susk}   S. Upadhyay,   S. K. Rai and B. P. Mandal,  J. Math. Phys.  {52}, {022301} (2011) 
 \bibitem{jog} S. D. Joglekar and A. Misra, Int. J. Mod. Phys. A 15, 1453 (2000)  


 \bibitem{sb1} S. Upadhyay and B. P. Mandal,  Eur. Phys. J.  {C 72},  2065 
(2012); Ann. Phys. {327}, 2885 (2012);    Mod. Phys. Lett.   {A 25}, {3347} (2010); Prog. Theor. Exp. Phys.   053B04,  (2014) 

\bibitem{smm} S. Upadhyay, M. K. Dwivedi and B. P. Mandal, Int. J. Mod. Phys. A 28, 1350033 (2013) 
\bibitem{rb} R. Banerjee and B. P. Mandal, Phys. Lett. B 488, 27 (2000)
\bibitem{bm1} R. Banerjee, B. Paul and S. Upadhyay,  Phys. Rev. D 88, 065
019 (2013)
\bibitem{rbs1} R. Banerjee  and S. Upadhyay, Phys. Lett. B 734, 369 (2014) 

\bibitem{sudd} S. Upadhyay and D. Das, Phys. Lett. B 733, 63 (2014) 
\bibitem{sud} S. Upadhyay, EPL  104, 61001  (2013);
   Phys. Lett. B 727, 293 (2013);  Ann. Phys. 340, 110 (2014) 

\bibitem{fs0} M. Faizal, S. Upadhyay and B. P. Mandal, Phys. Lett. B 738, 201 (2014) 
\bibitem{fs} M. Faizal, B. P. Mandal and S. Upadhyay, Phys. Lett. B 721, 159 (2013) 
\bibitem{lav} P. M. Lavrov and O. Lechtenfeld,  Phys. Lett. B 725,  382 (2013) 
\bibitem{ale} P. Y. Moshin and A. A. Reshetnyak, Nucl. Phys. B 888, 92 (2014)
\bibitem{ale1} P. Y. Moshin and A. A. Reshetnyak, Int. J. Mod. Phys. A 30, 1550021 (2015)    
\bibitem{b}M. Faizal and A. Higuchi, Phys. Rev. D78, 067502 (2008)
\bibitem{yo0} K. Yokoyama, Prog. Theor. Phys. 51, 1956 (1974) 
\bibitem{SU}S. Upadhyay,  EPL 105, 21001 (2014)
\bibitem{yok}  K. Yokoyama,
Prog. Theor. Phys. 59, 1699 (1978) 
\bibitem{yos} S. Upadhyay and B. P. Mandal, Prog. Theor. Exp. Phys.  053B04 (2014)
\bibitem{yoko}  K. Yokoyama,  Prog. Theor. Phys. 60, 1167 (1978);  Phys. Lett. B 79, 79   (1978) 

\bibitem{6}S. Upadhyay, Ann.   Phys.  344, 290 (2014)  
\bibitem{7}S. Upadhyay, Eur. Phys. J. C 74, 2737  (2014)  

 \bibitem{2}S. P. Miao, N. C. Tsamis, R. P. Woodard, J. Math. Phys. 50, 122502 (2009) 
 \bibitem{w2}S. P. Miao, N. C. Tsamis, and R. P. Woodard, J. Math. Phys. 51, 072503  (2010)
 \bibitem{w1}S. P. Miao, N. C. Tsamis, and R. P. Woodard, J. Math. Phys.
  52, 122301 (2011)
 \bibitem{4} 
 P. J. Mora, N. C. Tsamis, R. P. Woodard, J. Math. Phys. 52, 122301 (2011)  
 \bibitem{4a}  P. J. Mora, N. C. Tsamis, R. P. Woodard, J. Math. Phys.
 53, 2122502  (2012)
\bibitem{1}H. Bondi, M. G. J. van der Burg and A. W. K. Metzner, Proc. Roy. Soc. Lond. A 269, 21
(1962)
\bibitem{q1}
R. K. Sachs, Proc. Roy. Soc. Lond. A 270, 103 (1962)
\bibitem{q2}
A. Strominger, JHEP 1407, 152 (2014)
\bibitem{p}R. P. Woodard, arxiv:1506.04252


 \bibitem{1aaaa} R. Horvat, A. Ilakovac, J. Trampetic and J. You,  JHEP.  1112, 081 (2011) 
 \bibitem{2aaaa} H.  Grosse, H.  Steinacker and M.  Wohlgenannt,  JHEP. 0804, 023 (2008) 

 \bibitem{pqg1}M. Faizal,  J. Phys. A 44, 402001 (2011) 
\bibitem{pqg2}M. Faizal, Phys. Lett. B 705, 120 (2011) 
\bibitem{pqg3}M. Faizal, Mod. Phys. Lett. A 27, 1250075 (2012) 
\bibitem{pqg4a}J. W. Moffat, Phys. Lett. B 491, 345 (2000)   
\bibitem{pqg4}J. W. Moffat,  Phys. Lett. B 493,   142 (2000)
 
\end{thebibliography}
\end{document}